\documentclass[epj,final]{svjour}
\usepackage{latexsym}
\usepackage{bm}
\usepackage[dvips]{epsfig}
\usepackage[dvips]{graphicx}
\usepackage{bbm}

\newcommand{\sid}{\!} 
\newcommand{\sdf}{\,} 

\newcommand{\sff}{} 

\newcommand{\wideop}[1]{\,#1\,} 

\newcommand{\round}[1]{\left( #1 \right)}

\newcommand{\angular}[1]{\left[ #1 \right]}

\newcommand{\rd}{\mathrm{d}} 

\newcommand{\order}[1]{\mathcal{O}\round{#1}}

\newcommand{\e}{\mathrm{e}}

\newcommand{\sech}{\mathrm{sech\,}}

\newcommand{\multsum}[2]{\sum_{#1 \atop #2}}

\usepackage{amsbsy}
\usepackage{amssymb}

\newcommand{\pauli}[1]{\bm{\sigma}_{#1}} 
\newcommand{\id}{\mathbbm{1}_2}  
\newcommand{\mat}[1]{{\tens{#1}}} 
\newcommand{\blockmat}[1]{\bm{#1}} 
\newcommand{\tr}{\mathrm{Tr}}

\newcommand{\DA}{dynamical approximation }
\newcommand{\Tcval}{$T_c=(0.58912\pm0.00015)J$ }
\newcommand{\qtmex}{$\qt{}{m \neq 0}$-expansion}

\newcommand{\GS}[2]{\mathcal{S}_{#1}^{#2}} 
\newcommand{\pathintPsi}{\int \sid \mathcal{D}\;\Psi}
\newcommand{\Aeff}{\mathcal{A}_{\mathrm{eff}}} 
\newcommand{\muPF}{\mu_{\mathrm{{\tiny PF}}}} 

\newcommand{\Gi}[1]{\int_{#1}^G}  
\newcommand{\Giz}{\int_{\bm{\mathrm{z}}}^G}  
\newcommand{\Giy}{\int_{\bm{\mathrm{y}}}^G}  

\newcommand{\Gam}{\mat{\Gamma}} 
\newcommand{\Gamzero}{\mat{\Gamma}_{\mathrm{ stat}}} 
\newcommand{\Gzero}{\mat{G}_0} 
\newcommand{\Greg}{\mat{G}_{\mathrm{reg}}} 
\newcommand{\A}[2]{\mat{A}^{#1}_{#2}} 
\newcommand{\Vdyn}{\mat{V}_{\mathrm{dyn}}} 
\newcommand{\Vzero}{\mat{V}_{\mathrm{stat}}} 

\newcommand{\dP}{W}              
\newcommand{\dPS}{W_{\mathrm{stat}}}         
\newcommand{\dQ}{\Phi}         
\newcommand{\dQex}[1]{\Phi_{#1}} 
\newcommand{\dQextilde}[1]{\tilde{\Phi}_{#1}} 
\newcommand{\dQS}{\Phi_{\mathrm{stat}}}         

\newcommand{\Qb}{\bar{Q}} 

\newcommand{\Tcstat}{T_{c,{\mathrm{stat}}}}  

\newcommand{\wdiag}{w} 
\newcommand{\voff}{v} 

\newcommand{\qt}[2]{\tilde{q}^{#1}_{#2}}  
\newcommand{\qtsquare}[1]{{\tilde{q}_{#1}}^{\,\,\,2}}  
\newcommand{\qtstat}[1]{\qt{}{#1}}
\newcommand{\q}[1]{q_{#1}}  
\newcommand{\Hblock}[1]{\blockmat{v}_{#1}} 
\newcommand{\Hd}[2]{H_{#2}^{#1}}  
\newcommand{\Hr}{H_0}  
\newcommand{\z}[1]{z_{#1}}   
\newcommand{\y}[1]{y_{#1,0}}     
\newcommand{\yd}[3]{y_{#1,#2}^{#3}}  
\newcommand{\sus}[2]{\chi^{#1}_{#2}}  

\newcommand{\bb}{c}  

\newcommand{\C}{C}  

\newcommand{\expbb}[1]{\exp\left(#1\right)} 

\newcommand{\avdo}[2]{{\left[ #2 \right]}_{\mathrm{dis}}} 
\newcommand{\aveff}[2]{{\left\langle #2 \right\rangle}_{\mathrm{eff}}^{#1}} 
\newcommand{\avefftext}[2]{{\langle #2 \rangle}_{\mathrm{eff}}^{#1}} 

\newlength{\heightoffigures}
\begin{document}
\title{Dynamical solutions of a quantum Heisenberg spin glass model}
\author{M. Bechmann \and R. Oppermann 
}                     
%
%
\institute{Institut f\"ur Theoretische Physik und Astrophysik, Universit\"at W\"urzburg, D-97074 W\"urzburg, Federal Republic of Germany }
\date{Received: date / Revised version: date}
%
\abstract{
We consider quantum-dynamical phenomena in the $\mathrm{SU}(2)$, $S=1/2$ infinite-range quantum Heisenberg spin glass. 
For a fermionic generalization of the model we formulate generic dynamical self-consistency equations. 
Using the  Popov-Fedotov trick to eliminate contributions of the non-magnetic fermionic states
we study in particular the isotropic model variant on the spin space.
Two complementary approximation schemes are applied: one restricts the quantum spin dynamics to a manageable number of Matsubara frequencies while the other employs an expansion in terms of the dynamical local spin susceptibility. 
We accurately determine the critical temperature  $T_c$ of the spin glass to paramagnet transition. We find that the dynamical correlations cause an increase of $T_c$ by $2\%$ compared to the result obtained in the spin-static approximation. The specific heat $C(T)$ exhibits a pronounced cusp at $T_c$. Contradictory to other reports we do not observe a maximum in the $C(T)$-curve above $T_c$.

} 
\titlerunning{Dynamical solutions of a quantum Heisenberg spin glass model}
\maketitle
\section{Introduction}
  \label{intro}
  Theoretical models of spin glass systems, among which the best investigated is certainly the infinite-range Ising spin glass (SK model) \cite{sher_kirk_75}, are well known for their conceptual difficulties due to peculiarities of the ordered phase. Attempts of explicit solutions generally lead to severe numerical challenges. The infinite-range Heisenberg model, also known as the Quantum-SK model, is additionally complicated by the presence of quantum-dynamical correlations, even in the high-temperature phase. Realizations of Heisenberg spin glasses are discussed at length for example in the review by Binder and Young \cite{binder_young}, emphasizing the role played by different types of anisotropy in contrast to the isotropic model.

Theorists have been looking at the infinite-range Heisenberg spin glass model from different angles:
the quantum-dynamical self-consistency problem was first formulated by Bray and Moore in 1980 \cite{bray_moore_80} and the  corresponding TAP-equations have been derived by Sommers \cite{sommers_80,somm_usad_82}. Effects of external  fields and anisotropy  were also investigated \cite{gold_lai_HSG_90}. In these works explicit calculations relied on the spin-static approximation. Later, quantum-dynamical correlations  were taken into account, for instance by means of Quantum Monte Carlo simulations of the paramagnetic phase \cite{grem_roze_98_QMC} or by exact diagonalization of finite systems \cite{arra_rozenb_fin_T}. A different approach considers the SU(N)-generalization of the infinite range Heisenberg spin glass which can be solved exactly in the limit $N\to\infty$ \cite{sach_ye_93,geor_parc_sach_00,geor_parc_sach_01}.

There are physically fascinating and further reaching questions concerning the interplay of frustrated magnetism and other interactions. The competition between the Hei\-senberg spin glass interaction, Kondo-coupling, and transport, for instance, leads to non-Fermi liquid behavior \cite{burd_grem_geor_heavy_fermion_02}.
Despite the fact that such highly ambitious questions have been and continue to be addressed successfully, there remain many not yet resolved basic problems even in the infinite-range Heisenberg spin glass model alone.

This article is organized as follows. In sec. \ref{sec:general_theory} we introduce the fermionic model Hamiltonian and briefly sketch the dynamical spin glass decoupling procedure. We formulate general self-consistency equations for the fermionic $\mathrm{SU}(2)$, $S=1/2$ infinite range spin glass model.
Section \nolinebreak \ref{sec:isotropic_HSGS} concentrates on the isotropic Heisenberg  spin glass on the spin space which is contained in the general model as a special case. After a short digression to the spin-static approximation in sec. \ref{sec:static_approx} we apply in sec. \ref{sec:dynamical_solutions} the ``dynamical approximation'' which restricts the quantum spin dynamics to a feasible number of bosonic Matsubara frequencies.
We accurately locate the paramagnet to spin glass phase transition in sec. \ref{sec:Tc} and present our results for the specific heat in sec. \ref{sec:CV}. In order to support our results we additionally employ a perturbative expansion of the self-consistency equations in sec. \ref{sec:Vdyn_expansion}.  


\section{Formulation of the general dynamical self-consistency problem}
  \label{sec:general_theory}
  \subsection{Model definition and effective action}
    \label{sec:model_and_decoupling}
    We consider the grand-canonical Hamiltonian
\begin{equation}
\label{eq:hamiltonian}
  {\cal{K}}=\frac 1{2\sqrt{N}} \sum_{i\neq j,\nu}  J_{ij}^\nu S_i^\nu S_j^\nu \wideop{-} \sum_{i,\nu} h_\nu S_i^\nu \wideop{-} \mu \sum_{i,\sigma} a^\dagger_{i\sigma} a_{i\sigma},
\end{equation}
where the latin indices label the $N$ lattice sites and \linebreak $\nu=\{x,y,z\}$ denote the spatial directions. With the usual fermionic construction operators $a^\dagger$ and $a$ and the Pauli matrices $\pauli{\nu}$, the spin $1/2$ operators are represented by 
\begin{displaymath}
   S_i^\nu=\sum_{\sigma \sigma'}  a^\dagger_{i\sigma} \sigma^\nu_{\sigma\sigma'}a_{i\sigma'},
\end{displaymath}
where we dropped the conventional pre-factor $\hbar/2$ for convenience.
We assume quenched disorder among the magnetic couplings $J_{ij}^\nu=J_{ji}^\nu$ according to the symmetric Gaussian distribution 
\begin{equation}
\label{eqn:disorder_distribution}
P_\nu(J^\nu_{ij})=\frac{1}{\sqrt{2\pi}J_\nu}\exp\round{-\frac{\round{J_{ij}^\nu}^2}{2 {J_\nu}^{2}}}.
\end{equation}
The couplings in different spatial directions are completely uncorrelated. 

Although the Hamiltonian (\ref{eq:hamiltonian}) is defined on the Fock space it may readily be used to describe the corresponding model on the spin space, too. For this purpose one simply has to choose  the special imaginary and temperature-dependent chemical (Popov-Fedotov-) potential 
\begin{equation}
  \label{eq:mu_PF}
  \muPF=- i\pi T/2, 
\end{equation}
which effectuates exact cancellation of contributions of the superfluous non-magnetic (empty and doubly occupied) local states to the partition function \cite{popo_fedo_88}.

    We shall treat the general model (\ref{eq:hamiltonian}) within the framework of imaginary time Grassmann field theory, using the replica method  along with standard decoupling techniques \cite{bray_moore_80,op_mg_93}. Due to the infinite range of the magnetic interactions the system can be mapped easily onto a single-site problem by virtue of a site-global Hubbard-Stratonovich transformation. After a saddle point integration of the introduced decoupling fields the $n$-fold replicated and disorder averaged partition function is given by a Grassmann path integral,
\begin{equation}
  \label{eqn:Zn}
  \avdo{}{Z^n}=\pathintPsi \,\expbb{-\Aeff}.
\end{equation}
The resulting saddle point effective action reads 
\begin{eqnarray}
  \label{eq:effective_action}
  \Aeff&=&\int_\tau \bar{\Psi}_\tau \round{\round{\partial_\tau-\mu}\id\wideop{+}\sum_\nu h_\nu \pauli\nu}\Psi_\tau\\
&&\hspace{-0.5cm}\wideop{+}\sum_{a,b,\nu} J_\nu^2 \int_{\tau\tau'}\round{\frac 14\round{\Qb^{ab,\nu}_{\tau-\tau'}}^2\wideop{-}\frac 12 \GS{\tau}{a,\nu} \,\Qb^{ab,\nu}_{\tau-\tau'}\,\GS{\tau'}{b,\nu}}\nonumber,
\end{eqnarray}
where $a$ and $b$ are replica indices, $\Psi=\{\Psi_\uparrow,\Psi_\downarrow\}$ is a Grassmann spinor, $\GS{}{\nu}=\bar{\Psi}\pauli{\nu}\Psi$ denotes the Grassmann representation of a spin variable, and $\id$ is the $2\times 2$ identity matrix. 
The $\tau$-integrations extend from $0$ to $\beta=1/T$ and the real functions $\Qb^{ab,\nu}_{\tau-\tau'}$ satisfy the saddle point  conditions
\begin{equation}
  \label{eq:scQ}
  \Qb^{ab,\nu}_{\tau-\tau'}=\aveff{}{\GS{\tau}{a,\nu}\GS{\tau'}{b,\nu}}.
\end{equation}

Note that the Hamiltonian (\ref{eq:hamiltonian}) does not allow for qua-drupolar order, i.e. $\avefftext{}{\GS{\tau}{a,\nu}\GS{\tau'}{b,\mu}}\equiv 0$ for $\nu\neq\mu$. Our model differs slightly from the one with couplings of the type $ J_{ij}\sum_{i\neq j,\nu}S_i^\nu S_j^\nu$ that has been considered in other work \cite{bray_moore_80,sommers_80,somm_usad_82,gold_lai_HSG_90,grem_roze_98_QMC,arra_rozenb_fin_T}. In the paramagnetic phase, however, both model variants lead to  identical  effective actions and the results, particularly for the critical temperature, are thus directly comparable.

 \subsection{Self-consistency equations}
    \label{sec:seco_equations} 
    In the present article we consider properties of the model in the paramagnetic phase or in the spin glass phase close to the critical temperature such that a replica-symmetric treatment is sufficient. Hence, we introduce the components of a single time-independent spin glass order parameter,
\begin{equation}
  \label{eqn:def_q}
  \Qb^{a\neq b,\nu}_{\tau-\tau'}=\q{\nu}.   
\end{equation}

The time dependence of the problem merely resides in the replica-diagonal spin-spin correlations $\Qb^{aa,\nu}_{\tau-\tau'}$. We employ the Fourier decompositions of the time-dependent quantities,
\begin{eqnarray}
  \label{eqn:FTPsi}
  \Psi_\tau&=&T \sum_{l=\infty}^\infty  \Psi_l  \sff \e^{-i z_l \tau},\\
  \label{eqn:FTqt}
  \Qb^{aa,\nu}_{\tau-\tau'}&=&\sum_{m=-\infty}^\infty  \qt{\nu}{m} \sff \e^{-i\omega_m \round{\tau-\tau'}},
\end{eqnarray}
where  $z_l$ and $\omega_m $ denote fermionic and bosonic Matsubara frequencies, respectively.
The real Fourier coefficients $\qt{\nu}{m}=\qt{\nu}{-m}=\qt{}{\nu}\round{\omega_m}$ are the central quantities in our formulation of the theory.
These parameters are closely related to the local dynamical spin susceptibility:
\begin{equation}
  \label{eqn:chi_def}
  \sus{\nu}{m}=\sus{}{\nu}\round{\omega_m}=\beta \round{\qt{\nu}{m}\wideop{+}\q{\nu}\sff\delta_{m,0}}.
\end{equation}

By means of a second decoupling step the interacting  part of the effective action (\ref{eq:effective_action}) is rendered linear in the spin variables which permits the Gaussian integration of the Grassmann fields. 
For each spatial direction we introduce a replica-global decoupling field $\z{\nu}$ as well as a number of replica-local decoupling fields. The latter group into ``static'' fields $\y{\nu}$ and ``dynamical'' fields $\yd{\nu}{m\geq 1}{\pm}$ decoupling spin-spin interactions which are diagonal and off-diagonal in the fermionic frequency space, respectively.

This dynamical decoupling procedure leads to a system of non-interacting particles in the presence of an effective dynamical potential $\mat{V}$. In the space of the fermionic frequencies $z_l$, $\mat{V}$ takes the form of a non-diagonal Toeplitz-structured matrix defined by (for the sake of better readability we drop the replica indices from now)
\begin{equation}
  \label{eqn:V}
  \round{\mat{V}}_{l'l}=\left\{
    \begin{array}{cl}
      \Hblock{m},& l'=l+m, m>0,\\\vspace{-0.3cm}\\
      \Hblock{0},& l'=l,\\\vspace{-0.3cm}\\
      \Hblock{m}^\dagger,& l'=l-m, m>0,
    \end{array}
  \right.
\end{equation}
where the entries $\Hblock{m}$ themselves are $2\times 2$ matrices in spin space,
\begin{equation}
  \label{eqn:vblock}
  \Hblock{m}=\sum_\nu \sdf \pauli{\nu} \sff \Hd{\nu}{m}.
\end{equation}
Here the static effective magnetic fields are given by
\begin{equation}
  \label{eqn:static_H}
  \Hd{\nu}{m=0}=h_\nu + J_\nu\round{\sqrt{\q{\nu}}\;\z{\nu} \wideop{+} \sqrt{\qt{\nu}{0}-\q{\nu}}\;\y{\nu}},
\end{equation}
whereas the dynamical effective fields are complex and read
\begin{equation}
  \label{eqn:dynamical_H}
  \Hd{\nu}{m\geq 1}=J_\nu \sqrt{\frac{\qt{\nu}{m}}{2}}\sff\round{\yd{\nu}{m}{+}\wideop{+}i\sff\yd{\nu}{m}{-}}.
\end{equation}

As an auxiliary quantity we define the local one-particle Green's function in the presence of a particular configuration of these effective fields,
\begin{equation}
  \label{eqn:def_Gamma}
  \Gam=\round{\Gzero^{-1}\wideop{+}\mat{V}}^{-1},
\end{equation}
where 
\begin{equation}
  \label{eqn:G0_def}
        \round{\Gzero^{-1}}_{l'l}=\round{i z_l+\mu}\delta_{l',l}\sff\id
\end{equation}
is the inverse of the Green's function of the non-interacting system (i.e. $J_\nu=0$).

    We condense the notation by the use of a space saving abbreviation for the Gaussian integral operator,
\begin{equation}
  \label{eqn:GI}
  \Gi{x} f(x)=\frac{1}{\sqrt{2\pi}}\int_{-\infty}^{\infty}  \rd x \, \expbb{-\frac{x^2}{2}} f(x).
\end{equation}
Furthermore, we introduce the following shorthand notations for the multiple integrations over all occurring $z$-type and $y$-type decoupling fields:
\begin{eqnarray}
  \label{eqn:def_z_int}
    \Giz&=&\prod_\nu \Gi{\z{\nu}}\\
  \label{eqn:def_y_int}
    \Giy&=&\prod_\nu \round{\Gi{\y{\nu}}\prod_{m\geq 1}\Gi{\yd{\nu}{m}{+}}\Gi{\yd{\nu}{m}{-}}}
\end{eqnarray}

    Within the replica method, the disorder-averaged free energy per spin, $f$, is obtained by virtue of the relation
\begin{equation}
  \label{eqn:replica_trick}
  \beta f=-\lim_{n\to\infty} \frac{ \avdo{}{Z^n}-1}{n}
\end{equation}
which yields
\begin{equation}
  \label{eqn:free_energy}
  \beta f=-\sum_\nu \frac {\beta^2 {J_\nu}^2}{4}\round{\q{\nu}^2 -\sum_m\sdf\round{\qt{\nu}{m}}^2  }\wideop{-}\Giz\sdf \ln\dQ,
\end{equation}
where $\dQ$ is given by
\begin{equation}
  \label{eqn:def_Q}
  \dQ=\Giy \dP.
\end{equation}
The weight function
\begin{equation}
  \label{eqn:def_P}
  \dP=\det\round{\Gam^{-1}}/\det\round{\Greg^{-1}}
\end{equation}
results from the Gaussian integration of the Grassmann fields. For $\dP$ to be finite and meaningful a regularization of the determinant in eq. (\ref{eqn:def_P}) is required. In this work the simple choice 
\begin{equation}
        \label{eq:G_reg}        
        \round{\Greg^{-1}}_{l'l}=i z_l \sff \id
\end{equation} 
suffices.

By extremization of the free energy (\ref{eqn:free_energy}) with respect to the parameters $\q{\nu}$ and $\qt{\nu}{m}$ we finally derive the  self-consistency equations
\begin{eqnarray}
  \label{eqn:sc_q}
        \q{\nu}&=&\frac 1 {\beta^2}\Giz \frac 1 {\dQ^2} \round{\Giy \sdf\dP\sff\tr \;\A{\nu}{0}\Gam}^2,\\
  \label{eqn:sc_qtm}
        \qt{\nu}{m}&=&\frac 1 {\beta^2}\Giz \frac 1 \dQ \Giy \sdf\dP\times\nonumber \\
        &&\hspace{0cm} \round{ \round{ \tr\;\A{\nu}{m}\Gam}  ( \tr\;\A{\nu}{{-m}}\Gam) \wideop{-} \tr\;\A{\nu}{m}\Gam\A{\nu}{-m}\Gam}
\end{eqnarray}
employing the auxiliary matrices
\begin{equation}
  \label{eq:A_aux}
  \round{\A{\nu}{m}}_{l'l}=\pauli{\nu} \sff \delta_{l',l+m}.
\end{equation}

So far our formulation of the problem is rather general. With a suitable choice of the model parameters $J_\nu$, $h_\nu$, and $\mu$ eqs. (\ref{eqn:V})--(\ref{eqn:sc_qtm}) can be used to investigate the Heisenberg- and XY spin glasses or the Ising spin glass in a transversal field, both on the Fock space as well as on the spin space. Note that the direction-dependent distribution (\ref{eqn:disorder_distribution}) also allows for anisotropy. 

\section{The isotropic Heisenberg spin glass on the spin space}
  \label{sec:isotropic_HSGS}
  For the rest of this article we shall be concerned with the Heisenberg spin glass on the spin space. Therefore, the chemical potential will be fixed to $\mu=\muPF$ given by eq. (\ref{eq:mu_PF}). We only consider the isotropic model in the sense that the distribution of the magnetic couplings (\ref{eqn:disorder_distribution}) are independent of the spatial direction, i.e. $J_\nu\equiv J$, and consequently so are the spin-spin correlations, i.e. $\q{\nu}\equiv \q{}$ and $\qt{\nu}{m}\equiv \qt{}{m}$. Furthermore, we restrict ourselves to the case of zero external fields.

In anticipation of the following sections we state the important exact sum rule
\begin{equation}
  \label{eqn:sum_rule}
  \sum_{m=-\infty}^\infty \qt{}{m}=1
\end{equation}
which arises from eqs. (\ref{eq:scQ}) and (\ref{eqn:FTqt}) at equal times due to the absence of non-magnetic local states in the model on the spin space.

    \subsection{The spin-static approximation revisited}
      \label{sec:static_approx}
      As a starting point for the quantum-dynamical calculations in secs.  \ref{sec:dynamical_solutions} and \ref{sec:Vdyn_expansion} and as a simple but instructive special case of the eqs. (\ref{eqn:V}-\ref{eqn:sc_qtm}) we consider in this section the spin-static approximation \cite{op_mg_93}.

The effective potential matrix (\ref{eqn:V}) can be decomposed into a static and a dynamical part which are diagonal and non-diagonal in the fermionic frequency space, respectively:
\begin{equation}
  \label{eqn:V_splitting}
        \mat{V}=\Vzero\wideop{+}\Vdyn.
\end{equation}
According to the definition (\ref{eqn:V}) $\Vzero$ is a block-diagonal matrix composed of the static fields (\ref{eqn:static_H}) whereas $\Vdyn$ comprises the complex dynamical fields (\ref{eqn:dynamical_H}).

In the spin-static approximation the time dependence of the saddle point (\ref{eq:scQ}) is disregarded in the self-consistency problem. Consequently, the Fourier components $\qt{}{m\neq 0}$ are neglected in the construction of the effective potential and hence $\Vdyn=0$.

With the (frequency-diagonal) spin-static propagator matrix
\begin{equation}
  \label{eqn:Gamma_stat_def}
        \Gamzero=\round{\Gzero^{-1} \wideop{+}\Vzero}^{-1}
\end{equation}
the weight function (\ref{eqn:def_P}) can be calculated analytically resulting in
\begin{equation}
  \label{eqn:PS}
  \dPS=\frac 12 \cosh \round{\beta \Hr},
\end{equation}
where
\begin{equation}
  \label{eqn:H0_def}        
  \Hr=\sqrt{\sum_{\nu} \round{\Hd{\nu}{0}}^2}
\end{equation}
and the $\Hd{\nu}{0}$ are given by eq. (\ref{eqn:static_H}).
Since $\Vdyn=0$ the Gaussian integrations over the dynamical decoupling fields $\yd{\nu}{m\geq 1}{\pm}$ in eqs. (\ref{eqn:def_Q})--(\ref{eqn:sc_qtm}) become trivial. If we restrict this discussion to the paramagnetic phase (where the spin glass order parameter vanishes, i.e. $\q{}=0$) eq. (\ref{eqn:def_Q}) evaluates to 
\begin{equation}
  \label{eqn:Qstat}
  \dQS=\round{1+\bb^2}\exp\round{\frac{\bb^2}2}
\end{equation}
with the abbreviation $\bb=\beta J \sqrt{\qt{}{0}}$.

The dynamical approach made in this article facilitates the calculation of the dynamical saddle point components $\qt{}{m}$ even within the spin-static approximation. It can be seen easily that the first trace term in eq. (\ref{eqn:sc_qtm}) vanishes for $m\neq 0$ for a frequency-diagonal matrix $\Gam$. The second trace term, however, measures the overlap of two factors $\Gam$ that are displaced about $m$ elements along the diagonal which yields
\begin{eqnarray*}
  \frac 13 \sum_\nu  \tr \; \A{\nu}{m}\Gamzero\A{\nu}{-m}\Gamzero=&&\nonumber\\
  \label{eqn:Bm_static}
   &&\hspace{-3cm}  \sum_l\frac{2\round{i z_l+\mu}\round{i z_{l+m}+\mu}-\frac 23 \Hr^2}{\round{\round{i z_l+\mu}^2-\Hr^2}\round{\round{i z_{l+m}+\mu}^2-\Hr^2}}.
\end{eqnarray*}
Performing the fermionic Matsubara sum  and the angular integration of the $\y{\nu}$-fields (the $\z{\nu}$-integrations become trivial in the paramagnetic phase) finally leads to 
\begin{equation}
  \label{eqn:qtm_static}
  \qtstat{m}=\frac 1{3\dQS}\Gi{r}r^2\round{\frac{\bb r \sinh \round{\bb r}}{\bb^2 r^2+\pi^2 m^2}  \wideop{+}  \frac {\delta_{m,0}}2 \cosh \round{\bb r} }.
\end{equation}
For the static component this result constitutes a self-consistency equation with the explicit solution
\begin{equation}
  \label{eqn:qt0_static}
  \qtstat{0}=\frac {-3+\beta^2 J^2+\sqrt{9+30\beta^2J^2+\beta^4J^4}}{6 \beta^2J^2}.
\end{equation}
It is worth mentioning that the spin-static approximation in the shape of eq. (\ref{eqn:qtm_static}) exactly fulfills the sum rule (\ref{eqn:sum_rule}).

    \subsection{Dynamical approximations}
      \label{sec:dynamical_solutions}
      In order to study the role played by quantum-dynamical correlations we adopt the method of dynamical approximations that was introduced recently in the context of an itinerant spin glass model \cite{dyn_cpa_epjb}. 

In essence, this method systematically improves the spin-static approximation by successively taking into account the dynamical contributions to the effective potential $\mat{V}$ (\ref{eqn:V}). More precisely, in the so-called \DA  of order $M$  all Fourier components $\qt{}{m}$ with $m=\{0,..,M\}$ are kept in the self-consistency structure. The higher frequency components are set to zero in the construction of the effective potential. Thus, $\omega_M$ plays the role of a cut-off frequency for the dynamical self-interaction. Technically, at order $M$ the effective potential $\mat{V}$ is approximated by a band-diagonal matrix with band width $2(2M+1)$.

The main benefit of this approximation scheme is that all Gaussian integrations over the dynamical fields $\yd{\nu}{m>M}{\pm}$ become trivial which allows the numerical evaluation of the self-consistency equations (\ref{eqn:sc_q}) and (\ref{eqn:sc_qtm}) if the order $M$ is small enough.

The general strategy is to calculate a quantity within several dynamical approximations of increasing order $M$. Provided such a sequence of improved solutions reveals sufficient convergence properties the exact full dynamical result can be inferred by extrapolation to $M\to\infty$. A similar approximation technique on the discretized imaginary time space has been used earlier in the context of the Ising spin glass in a transverse field \cite{usad_buet_90b,gold_lai_90}.

In this article we present numerical solutions with $M$ ranging from $M=0$ (spin-static approximation, see sec. \ref{sec:static_approx}) to $M=4$. Up to $M=2$ the occurring integrations are performed by Gaussian quadrature. Due to the high dimensionality of the resulting integration problem for $M\geq 3$ we applied a combination of Gaussian quadrature and a Monte Carlo method for some less important angular integrations. The latter is the origin of the statistical errors in the numerical data presented in the following sections. 

The sum rule (\ref{eqn:sum_rule}) was derived for the fully dynamical system. Nevertheless, it was verified to be fulfilled in the cases $M=1,2$ by high precision numerical calculations. Although we have yet been unable to prove it analytically for $M>0$, we claim that  eq. (\ref{eqn:sum_rule}) holds exactly in any finite order of the dynamical approximation.

        \subsubsection{Solutions in the paramagnetic phase and determination of the critical temperature}
        \label{sec:Tc}
\begin{figure}
  \begin{center}
    \epsfig{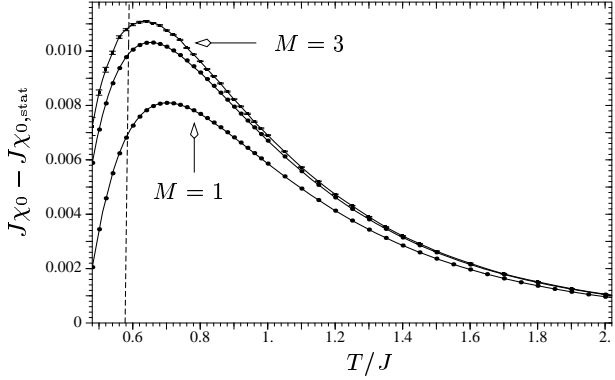}
    \caption{Paramagnetic solutions for the $\omega_m=0$ part of the local susceptibility $\chi_0=\beta \qt{}{0}$ in the dynamical approximations up to third order. Only deviations from the result in the spin-static approximation $\chi_{0,0}$ (eq. (\ref{eqn:qt0_static})) are shown. The dashed line represents $1-J\chi_{0,0}$;  according to eq. (\ref{eqn:Tc_condition}) the intersection points mark the respective approximation to the critical temperature. To the right of the dashed line the shown paramagnetic solutions are  unstable against spin glass order.}
    \label{fig:qt0}
  \end{center}
\end{figure}

\begin{figure}
  \begin{center}
    \epsfig{file=Tc.epsi,height=\heightoffigures}
    \caption{Sequence of critical temperatures obtained within dynamical approximations of orders $M=\{0,..,4\}$. The open diamond and the dashed line represent the spin-static  and the extrapolated full dynamical result, respectively. The arrows mark the critical temperatures obtained using the two lowest orders of the \qtmex of the functional $\dQ$ (\ref{eqn:def_Q}) discussed in sec. \ref{sec:Vdyn_expansion}. Inset: Extrapolation to $M=\infty$ (open star) having regard to the $1/M^3$-like convergence. The two thin lines indicate the statistical error.}
    \label{fig:Tc}
  \end{center}
\end{figure}

Our results for the zero-frequency part of the local susceptibility $\chi_0$ (\ref{eqn:chi_def}) obtained in the first three dynamical approximations are presented in fig. \ref{fig:qt0}. 
We find that the quantum-dynamical corrections to $\chi_0$ relative to the spin-static approximation are quantitatively remarkably small (note the small vertical scale in fig. \ref{fig:qt0}). This fact was already pointed out in \cite{grem_roze_98_QMC}. At high temperatures we observe quick convergence of this sequence of solutions meaning that the quantum dynamics of the model is described virtually exactly by taking into account only the effects of very few Matsubara frequencies. Naturally, as the temperature decreases the number of the relevant frequencies increases.

The equilibrium critical temperature of the paramagnet to spin glass phase transition can be determined by means of the simple relation
\begin{equation}
  \label{eqn:Tc_condition}
  J \,\chi_{0}\round{T_c}=1
\end{equation}
which can be shown to hold in any order of the dynamical approximation by expansion of eq. (\ref{eqn:sc_q}) in  powers of the order parameter $q$. Our solutions of eq. (\ref{eqn:Tc_condition}) with $M=\{0,..,4\}$ are presented in fig. \ref{fig:Tc}. 
From the structure of the self-consistency problem one expects a $M^{-3}$-like convergence of this sequence of $T_c$-approximants (see sec. \ref{sec:app_Vex}).
Extrapolation of the data to $M\to \infty$, i.e. to the full quantum-dynamical solution, yields \Tcval. This is an increase relative to the spin-static  result $\Tcstat =1/\sqrt{3} \,J$ by about $2\%$. Our result can also be compared \cite{compare_Tc} to values for $T_c$ obtained by means of quantum Monte Carlo simulations ($T_c=0.568\,J $) \cite{grem_roze_98_QMC} and exact diagonalization of finite systems ($T_c\approx 0.52 \,J$) \cite{arra_rozenb_fin_T}.

        \subsubsection{The specific heat}
        \label{sec:CV}
\begin{figure}
  \begin{center}
    \epsfig{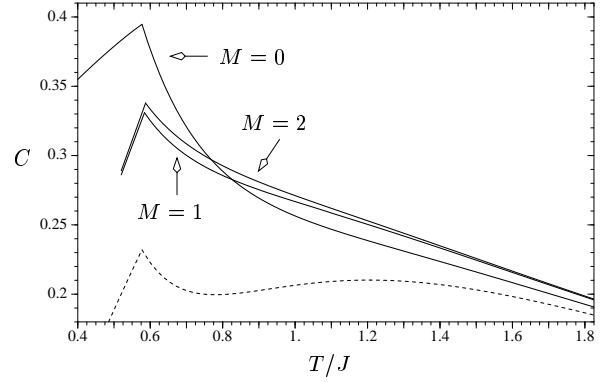}
    \caption{Specific heat according to eqs. (\ref{eq:specific_heat}) and (\ref{eq:internal_energy}) in the dynamical approximations with $M\leq 2$ (full lines). The cusp of the curves indicates the respective critical temperature $T_{c,M}$. Except for $M=0$, in the spin glass phase the results are correct close to $T_{c,M}$ only (see text). For comparison the ``conventional'' spin-static approximation, where the frequency sum in eq. (\ref{eq:internal_energy}) is restricted to the $m=0$ term, is also shown (dashed line).}
    \label{fig:specific_heat}
  \end{center}
\end{figure}

Using standard thermodynamic relations starting from the free energy (\ref{eqn:free_energy}) we derive a useful expression for the internal energy per site,
\begin{equation}
   \label{eq:internal_energy}
   U=\frac{3}{2} \beta J^2 \round{q^2\wideop{-}\sum_{m=-\infty}^\infty\sid\qtsquare{m}},
\end{equation}
from which the specific heat 
\begin{equation}
  \label{eq:specific_heat}
\C(T)=\frac{\rd U}{\rd T}
\end{equation}
can be  obtained by numerical evaluation of the temperature derivative.
Within the dynamical approximation of order $M$ the contributions to the frequency sum in eq. (\ref{eq:internal_energy}) with $|m|>M$ were calculated non-self-consistently from eq. (\ref{eqn:sc_qtm}) (or eq. (\ref{eqn:qtm_static}) in the spin-static case $M=0$).

In order to investigate the behavior of the specific heat for $T\lesssim T_c$ one needs solutions of eqs. (\ref{eqn:sc_q}) and (\ref{eqn:sc_qtm}) in the spin glass phase which are easily found over the whole temperature range in the spin-static approximation. 
For $M>0$, however, the full integration problem is hardly feasible. Hence, we restrict ourselves to temperatures sufficiently close to $T_c$ such that the self-consistency equations can well be approximated by expansions in powers of $T_c-T$. From eq. (\ref{eqn:sc_q}) we obtain for the spin glass order parameter the linear expression
\begin{equation}
  \label{eq:q_expansion}
  q=a\sff\round{T_c-T},\hspace{0.5cm} T\lesssim T_c,
\end{equation}
with the slope
\begin{equation}
  \label{eq:a_def}
  a=\frac 1J-\left.\frac {\rd}{\rd T} \qt{}{0}\right|_{T=T_c}.
\end{equation}
Expansion of eq. (\ref{eqn:sc_qtm}) in powers of $q$ yields the simplified self-consistency equation
\begin{equation}
  \label{eq:qtm_expansion}
  \qt{}{m}=\left. R_m\right|_{q=0}\wideop{+} c_m\sff q^2,\hspace{0.5cm} T\lesssim T_c,
\end{equation}
where $R_m$ symbolizes the right hand side of eq. (\ref{eqn:sc_qtm}) and the $c_m$ are well defined expansion coefficients that can be calculated numerically at $T=T_c$. 
By using the solutions of eqs. (\ref{eq:q_expansion}) and (\ref{eq:qtm_expansion}) in eq. (\ref{eq:internal_energy}) we obtain curves for the specific heat in the ordered phase that are correct at linear order of $T_c-T$. Note in particular that effects of Parisi replica symmetry breaking will change the results only in higher orders of $T=T_c$.

The resulting specific heat approximants with $M=\{0,.., 2\}$ are shown in fig. \ref{fig:specific_heat}. 
Due to the apparent quick convergence of this sequence of solutions we may safely draw qualitative conclusions for the limit $M\to\infty$. In the paramagnetic phase $C(T)$ monotonically increases as the temperature is lowered. Contrary to what was previously reported by other authors \cite{arra_rozenb_fin_T} in our results there is no indication of a broad maximum in the full dynamical $C(T)$-curve above $T_c$. Merely the ``conventional'' spin-static approximation which neglects the quantum dynamics altogether and omits all $m\neq0$ terms in the internal energy formula (\ref{eq:internal_energy}) has such a feature (fig. \ref{fig:specific_heat}).

    \subsection{Perturbative expansion of the self-consistency equations in  powers of the dynamical parameters $\qt{}{m\neq 0}$}
      \label{sec:Vdyn_expansion}
\begin{figure}
  \begin{center}
    \epsfig{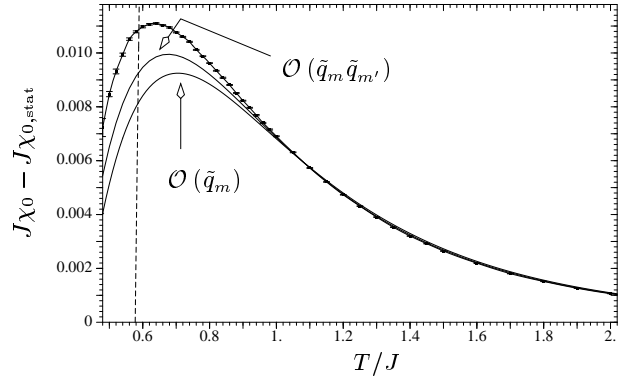}
    \caption{Paramagnetic solutions for $\chi_0=\beta \qt{}{0}$ (see fig. \ref{fig:qt0}) in the first two orders of the \qtmex  and in the third order \DA  (plot symbols) for comparison. Intersections with the dashed line mark the respective approximate critical temperatures which are indicated in fig. \ref{fig:Tc}.}
    \label{fig:qt0_Vdyn}
  \end{center}
\end{figure}

\begin{figure}
  \begin{center}
    \epsfig{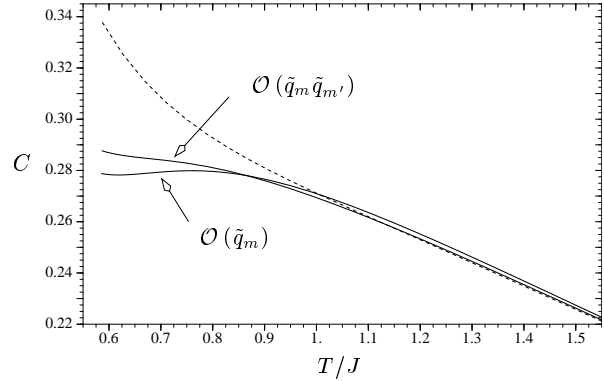}
    \caption{Specific heat in the first two orders of the $\qt{}{m}$-expansion and in the second order \DA (dashed line) for comparison (see fig.~\ref{fig:specific_heat}). Only solutions in the paramagnetic phase are shown.
}
    \label{fig:specific_heat_Vdyn}
  \end{center}
\end{figure}

\begin{figure}
  \begin{center}
    \epsfig{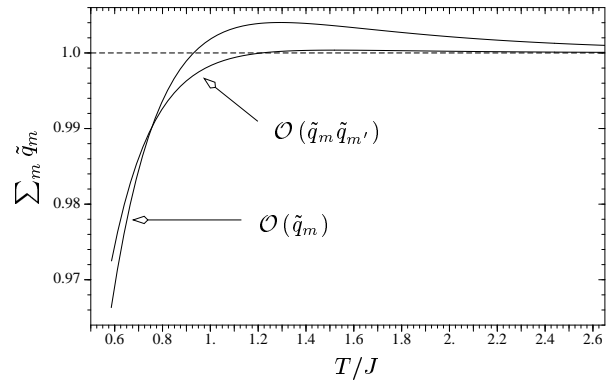}
    \caption{Check of the sum rule (\ref{eqn:sum_rule}) for the two lowest orders of the \qtmex indicating the quality of these approximations. Note that the higher order $\order{\qt{}{m} \qt{}{m'}}$ clearly does better for all temperatures.}
    \label{fig:sum_rule}
  \end{center}
\end{figure}

To consolidate our findings of sec. \ref{sec:dynamical_solutions} we additionally applied a completely different approximation technique to the self-consistency problem. Viewing the spin-static theory of sec. \ref{sec:static_approx} as a starting point we perturbatively expand the functional $\dQ$  (\ref{eqn:def_Q}) in  powers of the dynamical saddle point components $\qt{}{m\neq 0}$. Such an expansion is certainly justified at high temperatures where $ \qt{}{m\neq 0}\ll\qt{}{0}\lesssim 1$. 
With eq. (\ref{eqn:free_energy}) the \qtmex yields an approximate free energy which has to be extremized with respect to the parameters $\qt{}{m}$. Thus, we derive simplified self-consistency equations  (see \ref{sec:app_Vex}) that contain only integrations over the static decoupling fields $\y{\nu}$ and are therefore easily solved numerically.

The resulting solutions for the zero frequency susceptibility and the specific heat at the two lowest orders of the \qtmex $\order{\qt{}{m}}$ and $\order{\qt{}{m} \qt{}{m'}}$ are shown in fig. \ref{fig:qt0_Vdyn} and fig. \ref{fig:specific_heat_Vdyn}, respectively. At high temperatures the previous results  of the dynamical approximation discussed in sec. \ref{sec:dynamical_solutions} are reproduced very accurately. 
However, for $T/J\lesssim 1$ the results obtained  within the two different approximation schemes clearly differ from each other and apparently higher orders of the \qtmex become  important. 

From the numerical data presented in this section it is not obvious that the two sequences of approximations, the dynamical approximation with increasing $M$ on one hand, and the increasing orders of the \qtmex on the other hand, will finally converge to the same full dynamical solution as required. To prevent miss-interpre\-tations we check in fig. \ref{fig:sum_rule} how accurately the sum rule  (\ref{eqn:sum_rule}) is obeyed by the \qtmex at the two lowest orders. 
At high temperatures the sum rule is fulfilled almost exactly at order  $\order{\qt{}{m} \qt{}{m'}}$ reflecting the good quality of the approximation.
However, we observe  a substantial violation of the sum rule  for  $T/J\lesssim 1$ providing clear evidence that in this temperature regime this sequence of solutions is not well converged yet at the highest order considered in this article. 

Due to the small number of available orders we can not reliably extrapolate the results of the \qtmex to the full dynamical quantities. It can be seen in figs. \ref{fig:Tc} and \ref{fig:qt0_Vdyn}, however, that our results for the zero-frequency susceptibility presented in sec. \ref{sec:Tc}, and particularly our statement for $T_c$ are strongly supported by the present perturbative calculation. The specific heat, on the other hand, directly depends on the dynamical parameters $\qt{}{m\neq 0}$ and is therefore more sensible to the failure of the approximation to fulfill the sum rule (\ref{eqn:sum_rule}). We can nevertheless draw qualitative conclusions. The maximum obtained in the ``conventional'' spin-static approximation (see fig. \ref{fig:specific_heat}) being the zeroth order of the \qtmex is already very weak at order $\order{\qt{}{m}}$, and it is not present any more at order $\order{\qt{}{m} \qt{}{m'}}$. Hence, consistently with sec. \ref{sec:CV}, we claim that there is no maximum in the full dynamical $C(T)$-curve above $T_c$.

\section{Summary and conclusion}
  \label{sec:conclusion}
  In this article we applied standard techniques for many body systems to the fermionic $\mathrm{SU}(2)$, $S=1/2$ infinite range  spin glass model and formulated the general dynamical self-consistency problem for the spin-spin correlations in sec. \ref{sec:general_theory}. Using the Popov-Fedotov potential (\ref{eq:mu_PF}) to eliminate contributions of the non-magnetic local states we studied the isotropic Heisenberg spin glass on the spin space in sec. \ref{sec:isotropic_HSGS}. The results for the corresponding fermionic model on the Fock space will be published elsewhere.

In order to solve the highly coupled self-consistency equations and particularly to make the high-dimensional integration problem in eqs. (\ref{eqn:sc_q}) and (\ref{eqn:sc_qtm}) feasible we used two different systematic approximation schemes. 
The dynamical approximation of order $M$, on one hand, describes the quantum dynamics with a limited number of Matsubara frequencies but the corresponding saddle point components $\qt{}{|m|\leq M}$ are dealt with exactly (sec. \ref{sec:dynamical_solutions}). The perturbative \qtmex, on the other hand,  takes into account all frequencies but only a few powers of $\qt{}{m\neq 0}$ (sec. \ref{sec:Vdyn_expansion}). In this sense the two approximation  schemes are complementary to each other.

Both approaches yield a consistent picture for the zero frequency local spin susceptibility in the paramagnetic phase (figs. \ref{fig:qt0} and \ref{fig:qt0_Vdyn}). By extrapolation of the results in the dynamical approximation of the orders $M=\{0,..,4\}$ to $M\to\infty$ (fig. \ref{fig:Tc}) we obtained the full dynamical critical temperature  \Tcval which is about $2\%$ higher than the value in the spin-static approximation.  We also presented results for the specific heat $C(T)$ (figs. \ref{fig:specific_heat} and \ref{fig:specific_heat_Vdyn}).
In the framework of the dynamical approximation we perturbatively extended the calculations to the spin glass phase. We found a pronounced non-analyticity of $C(T)$ at $T_c$. We can not confirm the observation of a broad maximum in the $C(T)$-curve above $T_c$ which was reported by other authors \cite{arra_rozenb_fin_T}.  
In this point our results rather resemble those for a $\mathrm{SU}(N)$-generalization of the quantum Heisenberg spin glass in the limit $N\to\infty$ and for larger $S$ \cite{geor_parc_sach_00,geor_parc_sach_01}.

The methods discussed in this article are particularly useful to qualitatively and quantitatively describe the high temperature phases of disordered quantum systems. 
There are many open issues that can thus be addressed, e.g. real-frequency response functions, the behavior in a magnetic field, or questions concerning anisotropy.  
\\

This work was supported by the Deutsche Forschungs\-gemeinschaft under research project Op28/5--2 and by the SFB410. One of us (M. B.) also wishes to acknowledge  the scholarship granted by the University of W\"urzburg.

\begin{appendix}
\section*{Appendix}
\setcounter{section}{1}
    \subsection{Details of the perturbative \qtmex}
   \label{sec:app_Vex}
\begin{figure}
  \begin{center}
    \epsfig{file=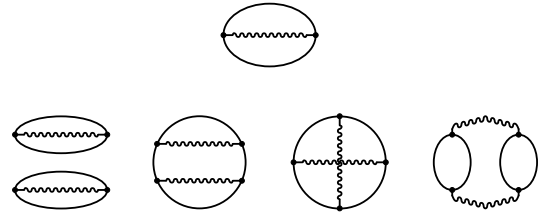,width=7cm}
    \caption{Diagrams contributing to the first two orders of the perturbative \qtmex of the functional $\dQ$ defined by eqs.  (\ref{eqn:def_Q}) and (\ref{eqn:P_expansion_prep}). Straight lines represent the spin-static propagator (\ref{eqn:Gamma_stat_def}), wavy lines symbolize contractions of the dynamical effective fields  (\ref{eqn:dynamical_H}).}
    \label{fig:diagrams}
  \end{center}
\end{figure}

As a consequence of the decomposition (\ref{eqn:V_splitting}) of the effective potential (\ref{eqn:V})  the weight function (\ref{eqn:def_P}) can be written as
\begin{equation}
  \label{eqn:P_expansion_prep}
        \dP=\dPS\sff\exp \tr \,\ln\round{\mat{1}\wideop{+} \Gamzero \Vdyn},
\end{equation}
where the spin-static quantities $\Gamzero$ and $\dPS$ are defined by eqs. (\ref{eqn:Gamma_stat_def}) and (\ref{eqn:Qstat}), respectively. 

In the high temperature limit the dynamical saddle point components vanish like $\qt{}{m\neq 0} \sim 1/T^2$ whereas the zero frequency component reaches unity, $\qt{}{0}\to 1$. Hence, at high temperatures the logarithm in eq. (\ref{eqn:P_expansion_prep}) can be expanded in powers of the matrix $\Vdyn$. The trace in eq. (\ref{eqn:P_expansion_prep}) and the Gaussian integrations over the dynamical decoupling fields $\yd{\nu}{m\geq 1}{\pm}$ in eq. (\ref{eqn:def_Q}) yield a representation of the functional
\begin{equation}
  \label{eqn:Q_expansion}
        \dQ=\dQS\wideop{+}\dQex{1}\wideop{+}\dQex{2}\wideop{+}\order{\qt{}{m}\qt{}{m'}\qt{}{m''}}
\end{equation}
in terms of contracted diagrams.

Figure \ref{fig:diagrams} displays all occurring diagrams up to the second non-trivial order. Evaluation of these diagrams yields the final results
\begin{equation}
  \label{eqn:Q1}
        \dQex{1}=\beta^2 J^2 \Gi{r} r^2 \sum_{m=1}^{\infty}\qt{}{m} \sff \frac{\bb\sff r \sinh \round{\bb \sff r}}{\bb^2 r^2+\pi^2 m^2} 
\end{equation}
and
\begin{eqnarray}
  \label{eqn:Q2}
        \dQex{2}&=&\beta^4 J^4 \Gi{r} r^2 \multsum{m=1}{m'\geq m}^{\infty} \qt{}{m}\qt{}{m'}
        \round{   \wdiag\angular{c\sff r,\round{\pi\sff m}^2 }\sff \delta_{m,m'} \right.\nonumber\\
                &&\hspace{2.6cm}\left. \wideop{+}\voff\angular{c\sff r,\round{\pi\sff m}^2, \round{\pi\sff m'}^2}}
\end{eqnarray}
with $\bb=\beta J \sqrt{\qt{}{0}}$ and the functions
\begin{eqnarray}
        \label{eqn:wdiag_def}
        \wdiag\angular{x,s}&=&\frac{\sech x}{\round{ x^2+s}^2}\round{  -\frac{3 s x^4-s^2 x^2+2s^3}{\round{x^2+s}\round{x^3+4 s \sff x}}\sinh\round{2x} \right.\nonumber \\
                   &&\hspace{-0.3cm}\left. +\frac{x^2-s+4}{8}\cosh\round{2x}+ \frac{11x^2-3s-4}{8}},\hspace{0.cm}
\end{eqnarray}
\begin{eqnarray}
        \label{eqn:voff_def}
        \voff\angular{x,s,t}&=&\frac{\sinh x}{\round{x^2+s}\round{x^2+t}}\left( \frac{x^2}{\cosh x \sff\sinh x} +\frac{2 s \sff x}{x^2+s} \right.\nonumber\\
                &&\hspace{0cm} \left.\wideop{+}\frac{2t\sff x}{x^2+t} -\frac{x\round{5x^2+s+t}\round{s+t}}{\round{x^2+s+t}-4 s \sff t}\right).
\end{eqnarray}
Expressions (\ref{eqn:Q_expansion})--(\ref{eqn:voff_def}) correctly reproduce the high temperature expansions of all quantities including order $\order{\beta^4}$.

In the context of the dynamical approximations discussed in sec. \ref{sec:dynamical_solutions}, particularly for the extrapolation to the full dynamical result of a quantity it is important to know how this quantity varies with the order $M$ for $M\to \infty$. It is sufficient to restrict the discussion to $\dQex{1}$ (\ref{eqn:Q1}). We write
\begin{equation}
  \label{eqn:Q1_split}
  \dQex{1}=\dQex{1,M}\wideop{+}\dQextilde{1,M},
\end{equation}
where for $\dQex{1,M}$ the sum in eq. (\ref{eqn:Q1}) is restricted to $m=\{1,..,M\}$  and $\dQextilde{1,M}$ contains the remaining high-frequency terms with $m>M$. Since $\qt{}{m}\sim m^{-2}$ for large $m$ (see eq. (\ref{eqn:qtm_static})) it is evident that $\dQextilde{1,M}\sim M^{-3}$ for large $M$. Thus, the functional $\dQex{}$ and consequently all derived quantities, e.g. the critical temperature, converge like $M^{-3}$.

\end{appendix}


\begin{thebibliography}{10}
\bibitem{compare_Tc}
{Note that the critical temperatures in references
  \cite{grem_roze_98_QMC,arra_rozenb_fin_T} have to be scaled by a factor of
  $4$ due to a slightly different definition of the spin operators.}
\end{thebibliography}
\end{document}